# Anomalous thermal conductivity in multiwalled carbon nanotubes with impurities and short-range order

V E Egorushkin, N V Melnikova, A N Ponomarev, A A Reshetnyak[1]

Institute of Strength Physics and Material Science, Siberian Branch of the Russian Academy of Sciences, 2/4, prospect Academicheskii, Tomsk, 634021, Russia

[1]E-mail: reshet@ispms.tsc.ru

**Abstract**. Low-temperature thermal conductivity and thermopower of multiwalled carbon nanotubes considered within a bundle of nanotubes are calculated taking into account multiple scattering of electrons on the atoms of impurities (like single carbon atoms) and short-range order regions arising due to a some distribution of the impurities. The calculations are realized on a base of the temperature diagrammatic Feynman techniques and the results of our research are in a good quantitative and qualitative agreement with the corresponding experimental data for multiwalled carbon nanotubes with diameter less or equal 10 nm at T<50K.

## 1. Introduction

Carbon nanotubes (CNTs) have unusual electron transport properties in a low temperature range (*T*<50*K*), which are similar to ones for amorphous metals and alloys with short-range order regions [1-3]. So, the low-temperature resistivity $\rho(T)$ of CNT decreases, but does not increase, when *T* rises [4]. As the consequence, the low-temperature conductivity σ(*T*) increases with grows of temperature approximately as $T^{1/2}$ [5]. In its turn, the thermal conductivity *k*(*T*) rises sharply in this temperature regime up to its maximum value and then decreases (see Fig. 1 [6]). At last, the thermopower *S*(*T*), being by more sensitive to a nature of electron transport, to be considered as a function of temperature is linear, positive, and well described by the Mott formula at *T*<50*K*. Near 50*K* the curve *S*(*T*) has a break or maximum (see Fig. 2 [6-8]) following by a change in the slope with respect to the *T* axis. At the same time, both the value of thermopower and that of *S*(*T*) slope, i.e. $\partial S(T)/\partial T$, have extraordinarily large magnitudes. Furthermore, in single-walled CNTs (SWNTs) exposed to air or oxygen, the positive values of thermopower becomes by negative ones for the SWNTs stripped of oxygen [8].

Conventionally, it is assumed that at *T>50K* the temperature behavior of the thermopower is determined by both the phononic subsystem and by the influence of defects on the symmetric electron-hole structure of CNTs. The latter influences on values of *S*(*T*), as well as of thermal conductivity $\rho(T)$, at T<50K unfortunately have been insufficiently understood at present [9,10]. As it was above-mentioned the amorphous metals and alloys with short-range order regions and CNTs have the same discriminating peculiarity to be concerned in the fact that the structure of all these systems includes the domains with short-range order. In particular, the defects of surface in CNTs are stipulated not only by possible declination from an ideal hexagonal structure, but also by the catalyst impurities, retained atoms, attached radicals (OH, CO and etc.), which change the configurations of valence and conduction bands.

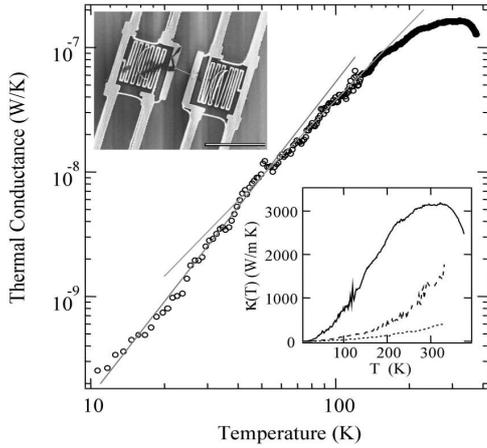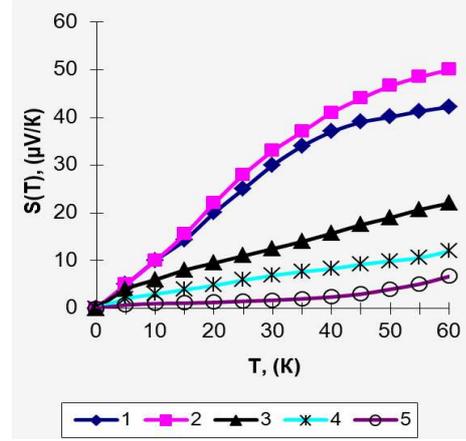

**Figure 1.** The thermal conductance of an individual MWNT of a diameter 14 nm see, Ref. [6]. Lower inset: Solid line represents $\chi(T)$ of an individual MWNT ($d$=14 nm). Broken and dotted lines represent small ($d$=80 nm) and large bundles ($d$=200 nm) of MWNTs, respectively. Upper inset: SEM image of the suspended islands with the individual MWNT. The scale bar represents 10 μm.

**Figure 2.** Temperature dependence of thermoelectric power of SCNT bundles to a diameter of $5 \cdot 10^{-6}$ m and a length of $10^{-3}$ m (curves 1,2 from Ref. [7]), bundles of CNTs with a diameter of $10^{-8}$ m and a length of $10^{-3}$ m (curve 3), MCNTs with a diameter of $10^{-8}$ m and a length of $2.5 \cdot 10^{-6}$ m (Curve 4 from Ref.[7]) and bundles of CNT diameter of $10^{-8}$ m and a length of $10^{-3}$ m (curve 5 in Ref.[6]).

Thus, the CNTs (not annealed to a purification) may be considered as "dirty" metallic systems where electrons are directly scattered by the structure defects: impurities and new chemical bonds forming the short-range order regions [11]. Within the such consideration we shall calculate relaxation time of electrons multiply elastically scattered on the structural inhomogeneities and then apply this result to obtain the corresponding contribution to electron thermal conductivity following to Wiedemann-Franz law. We shall also list the result of our calculation of the contributions to the low-temperature thermopower [11] where the one-electronic contribution is calculated by means of multiple elastic electronic scattering on the defects of the hexagonal CNT structure.

## 2. Electron relaxation time

Let us consider a solid (bundle of CNTs) with stochastically distributed defects such as catalyst impurities, retained atoms and so on (unconditionally of their origin). Then, we introduce microconcentrations (see, Ref.[12] for details) – the occupation numbers $c_s(\vec{R}_i)$ of the nodes denoted by the coordinates $\vec{R}_i = (x_i, y_i, z_i)$, $i=1,...,N$ by doped atoms of the type $s=1,2,...$. In absence of long-range order $c_s(\vec{R}_i) = c_s + \delta c_s(\vec{R}_i)$. Here $c_s$ is the macroconcentration of type $s$ atoms in the bundle of CNTs and $\delta c_s(\vec{R}_i)$ is the fluctuation of concentration. The average expectation-value of microconcentration with respect to configurations of the atoms is equal to, $\langle c_s(\vec{R}_i) \rangle = c_s$, whereas the average expectation-values for the fluctuation vanishes $\langle \delta c_s(\vec{R}_i) \rangle = 0$, and second order moment of $c_s(\vec{R}_i)$: $\langle \delta c_s(\vec{R}_i) \delta c_s(\vec{R}_j) \rangle$, $i,j=1,...,N$, will determine either new chemical bonds or new short-range order [12] in a system with defects. Its Fourier image $\langle \delta c_s(R_i) \delta c_s(R_j) \rangle \sim \langle |c_k|^2 \rangle$ determines the structure of short-range order region, so that, (for simplicity we restrict ourselves by the case of binary system, when $s=1$, $c_s = c$) $\langle |c_k|^2 \rangle = N^{-1} c(1-c) \sum_{i=1}^{N} \alpha_i \cos \vec{k}\vec{R}_i$, where $\alpha_i$ are the short-range order parameters and $\alpha_0 = 1$. The quantity $\langle |c_k|^2 \rangle_0 = N^{-1}c(1-c)$ determines an "ideal solid solution" for all $\alpha_{i\neq 0} = 0$ or presence of impurities under consideration at $c \leq 0,1,1$.

Then, we introduce a stochastic field of "new" atoms of unique type, for $s=1$

$$V(\vec{R}) = \sum_i U(\vec{R} - \vec{R}_i), \qquad (1)$$

where potential field $U(\vec{R} - \vec{R}_i)$ (till not completely determined) corresponds to one atom with $\vec{R}_i$. We will describe the electron interaction with new field $V(\vec{R})$ by the following Hamiltonian

$$H_{int} = \sum_i c(\vec{R}_i) \int \psi^+(\vec{R}) U(\vec{R} - \vec{R}_i) \psi(R) d\vec{R}, \qquad (2)$$

where $\psi^+, \psi$ are the field operators of electrons.

For simplicity, we suppose that $U(\vec{R} - \vec{R}_i) = U_0 \delta(\vec{R} - \vec{R}_i)$ because of the radius of action of the potential $U$ is enough small in comparison with the distance between atoms (sparse system). In the case under consideration, instead of the site potential we may pass to site $t$-matrices representing the amplitude of multiple scattering on the site. So $U_0$ may be considered as an effective potential $t_0$. Furthermore, we assume $U_0$ to be small what allows us to use only the first orders of perturbation theory series for amplitude of scattering with respect to $U_0$. For one-particle Green function (GF) $G$ the series looks as follows

$$G = G_0 + G_0 V G_0 + G_0 V G_0 V G_0 + ..., \qquad (3)$$

or, equivalently

$$G(\vec{r}, \vec{r}') = G_0(\vec{r}, \vec{r}') + \sum_i c(\vec{R}_i) \int G_0(\vec{r}, \vec{R}) U(\vec{R} - \vec{R}_i) G_0(\vec{R}, \vec{r}') d\vec{R} + \sum_{i,j} c(\vec{R}_i) c(\vec{R}_j)$$
$$\times \int\int G_0(\vec{r}, \vec{R}) U(\vec{R} - \vec{R}_i) G_0(\vec{R}, \vec{R}') U(\vec{R}' - \vec{R}_j) G_0(\vec{R}', \vec{r}') d\vec{R} d\vec{R}' + ... = G_0 + \sum_i G_i^{(1)} + \sum_{i,j} G_{i,j}^{(2)} + .... \qquad (4)$$

As to the rest terms of perturbation series one should be noted that after dividing in the each perturbation theory order one-node ($i = j = k...$) and two-node ($i \neq j \neq k...$) terms, which may be taken into account by including of $t$ – matrix instead of $U$, i.e. $U(\vec{R} - \vec{R}') = t_0 \delta(\vec{R} - \vec{R}')$. The rest more complicated scattering events should be neglected because the corresponding terms are proportional to $c^3$ or $<\delta c \delta c>^2$ being much less than $c^2$ or $<\delta c \delta c>$.

Now, we consider separately the first-order with respect to $U_0$ terms in the Eq.(4)

$$G_i^{(1)} = c(\vec{R}_i) \int G_0(\vec{r}, \vec{R}) U(\vec{R} - \vec{R}_i) G_0(\vec{R}, \vec{r}') d\vec{R}. \qquad (5)$$

In momentum representation, we have (with omitting the vector sign "$\rightarrow$" over $\vec{p}, \vec{R}, \vec{k}$)

$$G(r, R) = \int G(p) e^{ip(r-R)} dp, \qquad U(R - R') = \int U(k) e^{ik(R-R')} dk. \qquad (6)$$

Substituting expressions (6) into Eq. (5) we get the expression for propagator of one-particle GF

$$G_i^{(1)}(p, p') = c(R_i) e^{i(p-p')R_i} G_0(p) U(p - p') G_0(p'). \qquad (7)$$

Finally, averaging $G_i^{(1)}$ with respect to configurations $R$ and summarizing over $i$, we get

$$<G^{(1)}(p, p')> = c_s U_0 \delta(p - p') |G_0(p)|^2. \qquad (8)$$

Note, that in the Eq.(8) $U = U(0) = U_0$ without using the approximation $U(x) = U_0 \delta(x)$. In the latter case, $U(p - p') = U_0$. Expression (8) coincides for small $c$ with the analogous quantity $<G^{(1)}>$ for the case of "dirty" metals [12].

Let us turn to the next term in the Eq.(4)

$$G^{(2)}(r,r') = \quad \overset{\curvearrowright \quad \curvearrowright}{\underset{r \quad R \quad R' \quad r'}{\longrightarrow}}$$

or, equivalently, in analytical form,
$$G^{(2)}_{i,j} = \iint G_0(r-R)c(R_i)U(R-R_i)G_0(R-R')c(R_j)U(R'-R_j)G_0(R'-r')dRdR'.$$

In $p$ – representation after the change $p'' \leftrightarrow p'$ we have
$$\begin{aligned}G^{(2)}_{i,j}(p,p') &= c(R_i)c(R_j)G_0(p)G_0(p')\int U(p-p_1)G_0(p_1)U(p_1-p')e^{-i(p-p_1)R_i}e^{i(p'-p_1)R_j}dp_1 \\ &= G_0(p)G_0(p')\int U(p-p_1)G_0(p_1)U(p_1-p')c(R_i)e^{-i(p-p_1)R_i}c(R_j)e^{i(p'-p_1)R_j}dp_1\end{aligned} \quad (9)$$

Eq. (9) describes "double" scattering by one site and "single" scattering by two sites (the quotes assume the possibility of including $t$ –matrix that means the repeated multiple scattering by one site and by two sites, correspondently). It is useful estimate those terms in (9):

1) for $i = j \Rightarrow c^2(R_i) = c(R_i)$ and therefore
$$\begin{aligned}G^{(2)}_{i,i}(p,p') &= G_0(p)G_0(p')\int U(p-p_1)G_0(p_1)c(R_i)e^{-i(p-p')R_i}U(p_1-p')dp_1 \\ &= c(R_i)e^{-i(p-p')R_i}G_0(p)G_0(p')\int U(p_1-p')G_0(p_1)U(p-p_1)dp_1\end{aligned}. \quad (10)$$

After averaging of the propagator of GF $G^{(2)}_{i,i}(p,p')$ in (10) with respect to configurations of impurities and then summation over index $i$ [$\langle c(R_i)\rangle = c$; $\sum_i e^{-i(p-p')R_i} = \delta(p-p')$] we have
$$<G^{(2)}(p,p')>_{ii} = cU_0^2\delta(p-p')G_0^2(p)\int G_0(p_1)dp_1. \quad (11)$$

2) for $i \neq j$
$$G^{(2)}_{i \neq j} = G_0(p)G_0(p')\int U(p-p_1)G_0(p_1)U(p_1-p')c(R_i)e^{-i(p-p_1)R_i}c(R_j)e^{-i(p_1-p')R_j}dp_1.$$

Averaging and summation over $i,j$ lead to the representation
$$\begin{aligned}<G^{(2)}(p,p')>_{i\neq j} &= G_0(p)G_0(p')\int U(p-p_1)G_0(p_1)U(p_1-p')\Big\{\sum_{i\neq j}e^{-i(p-p_1)R_i}e^{-i(p_1-p')R_j}\{\langle c(R_i)c(R_j)\rangle \\ &\quad + c^2\}\Big\}dp_1 = U_0^2 G_0(p)\Big\{c^2 G_0^2(p)\delta(p-p') + G_0(p')\int \langle c(p-p_1)c(p_1-p')\rangle G_0(p_1)dp_1\Big\}\end{aligned}.(12)$$

One should be noted that the first term in the Eq.(12), $U_0^2 c^2 G_0^3(p)\delta(p-p')$, coincides with the corresponding term in the GF for the "dirty" metal [13].

In its turn, from the analysis of the second term,
$$U_0^2 G_0(p)G_0(p')\int <c(p-p_1)c(p_1-p')>G_0(p_1)dp_1, \quad (13)$$

under the suggestion that the electron scattering on the defects in disordered system to be elastic it follows $p = p'$ and the Eq. (13) takes the form
$$U_0^2 G_0^2(p)\int <|c(p-p_1)|^2>G_0(p_1)dp_1. \quad (14)$$

Collecting the resulting terms (8), (11), (12) together with use of the Eqs. (13), (14) we get
$$\begin{aligned}\langle G(p,p')\rangle &= G_0 + \langle G^{(1)}\rangle + \langle G^{(2)}\rangle_{ii} + \langle G^{(2)}\rangle_{i\neq j} = G_0 + cU_0\delta(p-p')|G_0(p)|^2 \\ &\quad + U_0^2 G_0^2(p)\delta(p-p')\Big[c\int G_0(p_1)dp_1 + c^2 G_0(p) + \int \langle|c(p-p_1)|^2\rangle G_0(p_1)dp_1.\Big]\end{aligned}, \quad (15)$$

The representation (15) permits one to find self-energy part $\Sigma$, electron relaxation time $\tau^{-1} = -2\mathrm{Im}\Sigma$ and properly the GF averaged over disorder. To this end, we start from the Dyson equation
$$G(p) = G_0(p) + G_0(p)\Sigma G(p), \quad (16)$$

or, in the first order with respect to $\Sigma$,
$$G(p) = G_0(p) + G_0^2(p)\Sigma, \tag{17}$$
From the Eqs. (15), (17) we may to derive the expression for $\Sigma$ as follows,
$$\Sigma = cU_0 + c^2U_0^2 G_0(p) + cU_0^2 \int G_0(p_1)dp_1 + U_0^2 \int \langle |c(p-p_1)|^2 \rangle G_0(p_1)dp_1. \tag{18}$$
The first two summands do not give contribution to the imaginary part of $\Sigma$, $\mathrm{Im}\,\Sigma$. However, they determine the contribution to a change of chemical potential (i.e. of the Fermi level), $\mu^* = \mu - cU_0$, and contributions of the following higher order infinitesimal.

Therefore, to calculate the electron relaxation time $\tau$ we have to consider two last terms in the Eq. (18). As the result for $\mathrm{Im}\,\Sigma = \mathrm{Im}\,\Sigma_1^{(2)} + \mathrm{Im}\,\Sigma_2^{(2)}$, we have, firstly,
$$\Sigma_1^{(2)} = cU_0^2 \int G_0(p_1)dp_1.$$
Pass to integrating with respect to scalar $\xi$, $\xi = p^2/2m - \mu$, and omitting the factor $(2\pi)^{-3}$ we get
$$\int G_0(p_1)d\vec{p}_1 = \frac{\nu_0}{4\pi}\int_{|n|=1}d\vec{n}\int d\xi G_0(\xi) = \nu_0 \int \frac{d\xi}{\varepsilon - \xi + i0\,\mathrm{sign}\,\varepsilon} = -i\pi\nu_0\,\mathrm{sign}\,\varepsilon,$$
with angle vector $\vec{n}$. Thus, we obtain for $\Sigma_1^{(2)}$ the expression,
$$\Sigma_1^{(2)} = -icU_0^2\pi\nu_0\,\mathrm{sign}\,\varepsilon = -\frac{i}{2\tau_{imp}}\,\mathrm{sign}\,\varepsilon, \tag{19}$$
where $\nu_0$ is the density of electron states on the Fermi level, and
$$\left(\tau_{imp}\right)^{-1} = 2\pi cU_0^2\nu_0. \tag{20}$$
The last expression coincides with the corresponding expression for electron relaxation time calculated within the theory of "dirty" metals with impurities [13, 14].

Secondly, for the quantity $\Sigma_2^{(2)}$ from (18), $\Sigma_2^{(2)} = U_0^2 \int \langle |c(p-p_1)|^2 \rangle G_0(p_1)dp_1$. and validity of the formula $\langle |c(k)|^2 \rangle = \frac{c(1-c)}{N}\sum_{i=0}^N \alpha_i \cos\vec{k}\vec{R}_i$ (see formula before Eq.(1)) [12] we derive,
$$\Sigma_2^{(2)} = U_0^2 \frac{c(1-c)}{N}\sum_{i=0}^N \alpha_i \int_0^\infty \frac{\cos(\vec{p}-\vec{p}_1)\vec{R}_i d\vec{p}_1}{\varepsilon - \xi(p_1) + i0\,\mathrm{sign}\,\varepsilon}.$$
The standard Euler formula for $\cos y$, $y = (\vec{p}-\vec{p}_1)\vec{R}$, $\cos y = \frac{1}{2}\left(e^{iy_i} + e^{-iy_i}\right)$ leads to the representation
$$\int_0^\infty \frac{e^{\pm i\vec{p}_1\vec{R}_i}R_i d\vec{p}_1}{\varepsilon - \xi(p_1) + i0\,\mathrm{sign}\,\varepsilon} = \frac{4\pi}{R}\int_0^\infty \frac{\sin(p_1 R_i)p_1 dp_1}{\varepsilon - \xi(p_1) + i0\,\mathrm{sign}\,\varepsilon}. \tag{21}$$
Therefore, $\Sigma_2^{(2)}$ has the form
$$\Sigma_2^{(2)} = -\frac{4\pi^2 U_0^2 mc(1-c)}{N}\sum_{i=0}^N \alpha_i \frac{\cos(\vec{p}\vec{R}_i)}{R_i}e^{ixR_i\,\mathrm{sign}\,\varepsilon}, \tag{22}$$
where we have introduced the notation for variable $x$, $x = \sqrt{2m(\varepsilon + \varepsilon_F + i0\,\mathrm{sign}\,\varepsilon)}$

Substituting expressions (19) and (22) into Eq. (18) we obtain the self-energy part of GF,
$$\Sigma(p,\varepsilon) = -\frac{i\sin\varepsilon}{2\tau_{np}} - \frac{4\pi^2 U_0^2 mc(1-c)}{N}\sum_{i=0}^N \alpha_i \frac{\cos(\vec{p}\vec{R}_i)}{R_i}e^{ixR_i\,\mathrm{sign}\,\varepsilon}. \tag{23}$$

It is easily to extract the real and imaginary parts of $\Sigma(p,\varepsilon)$ where the former term will correspond to the renormalization of the chemical potential $\mu \to \mu^* = \mu - \Sigma_{real}$ and, the latter one for small momentum of electrons: $|p| \to 0$ takes the form

$$\mathrm{Im}\Sigma(p,\varepsilon)_{|p|\to 0} = \mathrm{Im}\Sigma(\varepsilon) = -i\left(\frac{1}{2\tau_{np}} + \frac{4\pi^2 U_0^2 mc(1-c)}{N}\sum_{i=0}^{N}\alpha_i \frac{\sin xR_i}{R_i}\right) sign\varepsilon .$$

Therefore, for small electron energy $\varepsilon$ we have:

$$\mathrm{Im}\Sigma(\varepsilon) \approx -\tfrac{i}{2}(sign\,\varepsilon)\,\tau^{-1}, \tag{24}$$

where,

$$\frac{1}{2\tau} = \frac{1}{2\tau_{imp}} + 4\pi^2 U_0^2 mc(1-c)\frac{x}{N}\sum_{i=1}^{N}\alpha_i . \tag{25}$$

The parameter of short-range order $\alpha_i$ in (25) is negative in the first coordinate sphere ($\alpha_1 < 0$) and positive in the second one ($\alpha_2 > 0$). It is important that the coordinate sphere parameters $\alpha_i$ may change a sign in dependence of atoms which generate the pairs. The first term in (25), $\tau_{imp}^{-1}$ represents the inverse relaxation time of electrons scattered by impurities. The second one corresponds to the inverse relaxation time of electron scattered by structural inhomogeneities of short-range order type.

At last, the one-electron GF averaged with respect to configurations in "dirty" metallic CNT has the form (see, for details Refs.[12-14]):

$$\langle G(\varepsilon,p)\rangle = \left(\varepsilon - \xi(p) + \tfrac{i}{2}\tau^{-1}(p) sign\,\varepsilon\right)^{-1}, \tag{26}$$

where $\xi(p) = p^2/2m - \mu^*$, $\vec{p}$ and $m$ – momentum and mass of electron, $\mu^*$ is the renormalized electron chemical potential. We shall use expressions (25) and (26) to calculate the contribution of multiple elastic electronic scattering by the short-range order regions.

### 3. Thermal conductivity

As far as we consider model CNTs representing at low-temperature region the metallic systems with elastic electron scattering we may use the Wiedemann - Franz law to determine a thermal conductivity

$$\chi(T) = \tfrac{1}{3}\left(e^{-1}\pi k_B\right)^2 T\sigma(T). \tag{27}$$

where $k_B$, $e$, $n$ are respectively the Boltzman constant, charge of the electron and concentration of electrons with $\sigma(T) = m^{-1}e^2\tau n$ being by Drude conductivity. Because of the electron relaxation time is given by the Eq. (25), we have for the first coordination sphere, after the following substitution $\varepsilon \to \varepsilon_l = \pi T(2l+1)$ (for vanishing integer $l$)

$$\sigma(T) = \frac{ne^2}{4\pi^2 cU_0^2 mv_0}\frac{1}{1-BT^{1/2}}, \tag{28}$$

where $B = (v_0 N)^{-1} 2\sqrt{2\pi}(1-c)m^{3/2}$. Finally, we obtain

$$\chi(T) = \frac{k_B^2 n}{12cU_0^2 mv_0}\frac{T}{1-BT^{1/2}}, \tag{29}$$

It is useful to estimate the above contribution to thermal conductivity of multi-walled CNTs. Doing so, we have for the concentration of electrons in CNT, $n=10^{28}$-$10^{29}$m$^{-3}$; $e=1,6\cdot 10^{-19}$C; $m= 9,1\cdot 10^{-31}$ kg; $v_0=1$eV$^{-1}=0,625\cdot 10^{19}$ J$^{-1}$; $c=0,1$; $U_0^2 = (0,05\text{eV})^2 = 5\cdot 10^{-4}\cdot(1,6\cdot 10^{-19})^2$ J$^2$. Hence, we get

$\left(4\pi^2 c U_0^2 m v_0\right)^{-1} n e^2 \hbar = 7,1 \cdot 10^6 \, S \cdot m^{-1}$ and $B = (N p_0)^{-1} 4\pi^2 \sqrt{2\pi}(1-c) m^{3/2} k_B^{1/2} = 1,4 K^{-1/2}$ (with restored dependence of Plank constant $\hbar$ and momentum on the Fermi level $p_0$). Therefore, the estimation holds

$$\left(12 c U_0^2 m v_0\right)^{-1} k_B^2 n \hbar \approx 4,6 \, W \cdot K^{-2} \cdot m^{-1}, \tag{30}$$

that permits one to derive for $T \sim 50K$ the value for thermal conductivity, $\chi(T) \propto -26 W \cdot K^{-1} \cdot m^{-1}$. The last result is to be in a good qualitative and quantitative agreement with the experimental data [6] for multi-walled CNTs with impurities with diameter less or equal 10 nm and we present on Figure 3 the low-temperature dependence of the calculated thermal conductivity.

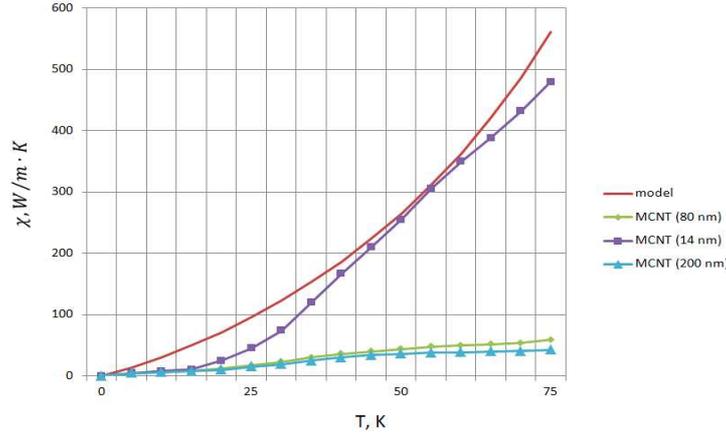

**Figure 3.** Thermal conductivity of a model and single MCNT with diameter 200 нм (curve 1), 80 нм (curve 2) and 14 нм (curve 3). (Obtained from the experimental data in Ref.[6]). Theoretical temperature dependence of the input of the multiple elastic scattering of the electrons on the short-range ordered regions in $\chi(T)$ (curve 4 as red type).

## 4. Thermopower

In our recent paper [11] we have proposed a theory of low-temperature thermopower in CNTs. On a base of the Mott formula

$$S(T) = \tfrac{1}{3}|e|^{-1} k_B^2 \pi^2 T \left(\tau \partial \tau^{-1} / \partial \varepsilon\right)_{\varepsilon = \varepsilon_F}, \tag{31}$$

we have calculated the corresponding one-particle contribution to thermopower in the form,

$$S(T) = -(18|e|N)^{-1} k_B^2 \pi m (1-c) \sum_i \alpha_i R_i^2 T, \quad R_0 = 2\pi (k_0)^{-1}, \tag{32}$$

where $R_i$, $\alpha_i$ are the radius of coordinate sphere and the parameters of short-range order, which includes the coordinate numbers [14] and $k_0$ is the wave-vector on Fermi level.

From (33) we would like to make the following conclusions:
- At low $T$ the magnitude and curve slope of $S(T)$ significantly depend on the size of inhomogeneity regions and do not determined by the parameters of electronic structure.
- The change of thermopower sign for MCNTs in the oxygenated and deoxygenated states related to the geometry of a new structure of MCNTs, so that for $\alpha_i > 0$ we have $S(T) < 0$, and vice versa ([10]).
- The value of thermopower correlates to the size of samples of CNTs (their diameter through $R_i$).
- With grows of the impurity concentration $c$ the magnitude $|S(T)|$ decreases in accordance with [10].
- The break or maximum of $S(T)$ around 50K may be due to forming of a stable region of structural inhomogeneities in CNTs (the so-called "mesoscopic first order phase transition" [6]), which depends on the change of number of the parameters $\alpha_i$ in eq. (32) part of which can be positive or negative in the neighborhood of a critical point $T_0$, hence changing the form of $S(T)$.

The estimation of one-electron contribution to thermopower described by the Eq. (32) permits to derive that $S(T) \propto -2 \cdot 10^{10} N^{-1}(1-c) \sum_i \alpha_i R_i^2 T$. For $N^{-1}(1-c) \sum_i \alpha_i R_i^2 \propto -0.5 \cdot 10^{-16}$ m$^2$, $\alpha_i R_i^2 < 0$, $\alpha_i \propto 10^{-2}$, $R_i \propto 10^{-7}$ m, $S(T) \propto T$ mkV/K. Therefore, for $T=10K$ we have $S(T) \sim 10$ mkV/K. This result is in a good agreement with the experimental data [6, 7] (see, for details Ref. [11]).

## 5. Conclusion

We have suggested the new model of multi-walled CNT with impurities within bundle of CNTs by means of introducing special distribution of the atoms of impurities and short-range order regions arising due to those impurities and defects of hexagonal lattice structure. The results of our first-principle research of the contributions of multiple elastic electron scattering on the impurities and short-range order regions in the framework of above model of CNT and temperature Green functions approach to thermal conductivity and thermopower are presented by the Eqs.(29),(32) and shown to be in a good quantitative and qualitative agreement with the corresponding experimental data for T<50K for multiwalled carbon nanotubes with diameter less or equal 10 nm. Hence, our model may describe unusual low-temperature behaviour of electron transport properties of metallized carbon CNTs.

In order to construct the complete theory of electron transport in CNTs at low temperatures we have to take into account inelastic electron-electron interaction (that is especially important for the electron resistance) and its interference with multiple elastic electron scattering on the defects. From the other hand, the enlargement of the results of temperature dependence of thermal conductivity and thermopower into region of $T>50K$ require to turn on the phononic subsystem of CNT into consideration. The resolution of the above two points are within scopes of our research.

**Acknowledgement**
The work is supported in part by the RFBR (Project № 09-02-992). The authors thanks the organizers of the DubnaNANO2010 Conference for support and hospitality.

**References**
[1]  Mizutani U 1993 *Phys. Stat. Sol.* (*b*) **176** pp. 9-30
[2]  Gyunterodt G I and Bek G (eds.) 1986 *Metallic Glasses, Vol. 2: Atomic Structure and Dynamics, Electronic Structure, and Magnetic Properties* (Mir, Moscow)
[3]  Zallen R 1998 *The physics of amorphous solids* (Wiley Classic Library Edition)
[4]  Jang J W, Lee D K, Lee C E, Lee T J, Lee C J, et al. 2002 *Solid State Comm.* **124** p.147
[5]  Postma H W Ch et al 2000 *Phys. Rev.* B **62** R10653 - R10656
[6]  Kim P, Shi, McEuen P L 2001 *Phys. Rev. Lett.* **87** No 21 215502
[7]  Kong, W.J. Lu, L. H W Zhu, H.W. Wei, B.Q. and Wu, D.H. 2005 *J. Phys. Cond. Matt.*, **17**, 1923-1928
[8]  Bradley K, Jhi S-H, Collins Ph G, Hone J, et al. 2000 *Phys. Rev. Lett.* **85** No 20 4361-4364
[9]  Grigorian L, Sumanasekera G U, et al. 1999 *Phys. Rev.* B **60** No 16 R11309-R11312
[10] Avouris Ph, Martel R, et al. 2000 *Electrical properties of carbon nanotubes: spectroscopy, localization and electrical breakdown. Science and Application of Nanoubes* ed. by Tomanek and Enbody (Kluwer Academic/Plenum Publishers, New York) pp 223–237
[11] Egorushkin V E, Melnikova N V, Ponomarev A N 2009 *Russ. Phys. J.* **52** № 3 252–264
[12] Levitov L S and Shitov A V 2003 *Green's Functions. Problems and Solutions*, 2nd ed. (Fizmatlit, Moscow), in Russian
[13] Abrikosov A A, Gorkov L P, and Dzyaloshinski I E 1963 *Methods of Quantum Field Theory in Statistical Physics* (Dover, New York)
[14] Iveronova V I and Katsnelson A A 1977 *Short-range Order in Solid Solutions* (Nauka, Moscow)